\documentclass[a4paper,12pt,preprint,amsmath,showpacs,
nofootinbib,superscriptaddress]{revtex4}
\usepackage[hypertex]{hyperref}
\usepackage{graphicx}
\usepackage{epstopdf}
\usepackage{bm}
\usepackage{epsfig}
\usepackage{graphics}
\usepackage{xspace}
\usepackage{amsfonts}

\def\Slash#1{{#1\!\!\!\slash}}

\newcommand{\nn}{\nonumber}

\newcommand{\ep}{\epsilon}
\newcommand{\al}{\alpha}
\newcommand{\be}{\beta}
\newcommand{\ga}{\gamma}
\newcommand{\Ga}{\Gamma}
\newcommand{\la}{\lambda}
\newcommand{\de}{\delta}
\newcommand{\De}{\Delta}
\newcommand{\bea}{\begin{eqnarray}}
\newcommand{\eea}{\end{eqnarray}}

\begin{document}
%\setlength\baselineskip{17pt}

%%%%%%%%%%%%%%%%%%%%%%%%%%%%%%%%%%%%%%%%%%
%Define Title, Author, Address, Preprint#

\title{ Search for the signal of monotop production at the early LHC }

\vspace*{1cm}

\author{Jian Wang}
\affiliation{Department of Physics and State Key
Laboratory of Nuclear Physics and Technology, Peking
University, Beijing, 100871, China}

\author{Chong Sheng Li\footnote{Electronic
address: csli@pku.edu.cn}}
\affiliation{Department of Physics and State Key
Laboratory of Nuclear Physics and Technology, Peking
University, Beijing, 100871, China}
\affiliation{Center for High Energy Physics, Peking
University, Beijing, 100871, China}

\author{Ding Yu Shao}
\affiliation{Department of Physics and State Key
Laboratory of Nuclear Physics and Technology, Peking
University, Beijing, 100871, China}

\author{Hao Zhang}
\affiliation{Department of Physics and State Key
Laboratory of Nuclear Physics and Technology, Peking
University, Beijing, 100871, China}

%\date{\today\\ \vspace{1cm} }

%%%%%%%%%%%%%%%%%%%%%%%%%%%%%%%%%%%%%%%%%%
%Create the title page

\begin{abstract}
 \vspace*{0.3cm}
We investigate the potential of the early LHC to discover the signal of monotops,
which can be decay products of some resonances in models such as
R-parity violating SUSY or SU(5), etc.
We show how to constrain the parameter space of the models by the present data of
$Z$ boson hadronic decay branching ratio, $K^0-\overline{K^0}$ mixing and dijet productions at the LHC.
Then, we study the various cuts imposed on the events, reconstructed from the hadronic final states,
to suppress backgrounds and increase the significance in detail.
And we find that in the hadronic mode the information from the missing transverse energy and reconstructed resonance mass
distributions can be used to specify the masses of the resonance and the missing particle.
Finally, we study the sensitivities to the parameters at the LHC with $\sqrt{s}$=7 TeV and
an integrated luminosity of $1~{\rm fb}^{-1}$ in detail.
Our results show that the early LHC may detect this signal at 5$\sigma$ level for
some regions of the parameter space allowed by the current data.

\end{abstract}


\maketitle
\newpage
%%%%%%%%%%%%%%%%%%%%%%%%%%%%%%%%%%%%%%%%%%
%Main body of the paper

%\tableofcontents

%\newpage
\section{Introduction}
\label{sec:1}
The main tasks of the Large Hadron Collider (LHC) are to answer the fundamental questions in particle physics:
whether the Higgs boson exist or not.
And are there new physics beyond standard model (SM) such as supersymmetry (SUSY), extra dimension, etc, at the TeV scale?
Generally, it is believed that top quark may have strong connections with new physics
due to its large mass close to the scale of electroweak symmetry breaking.
The production topologies of top quark pair production with or without missing
transverse energy $\Slash{E}_{T}$ have been extensively investigated \cite{Alvarez:2011hi,Haisch:2011up,Berger:2011hn,Cao:2010nw,
Degrande:2010kt,Battaglia:2010xq,Cao:2010zb,Alwall:2010jc,Han:2008gy,Barger:2006hm}.
However the topology of a single top and $\Slash{E}_{T}$, which is so-called monotop \cite{Andrea:2011ws},
has only been discussed recently \cite{Kamenik:2011nb,Dong:2011rh}.
This signal is absent in the SM and occur in models such as
R-parity violating SUSY and SU(5) as decay products of resonance production of some particles.
In R-parity violating SUSY \cite{Barbier:2004ez}, a stop can be produced by the fusion
of two down-type anti-quarks through the Yukawa-like trilinear interaction
$\lambda_{ijk}^{''}U_i^cD_j^cD_k^c$, where $U_i,D_i$ are left-handed chiral superfields
and the superscript $c$ denotes the charge conjugate,
and then the stop decays into a top quark and a neutralino which
could not be detected at the collider.
In the SU(5) model \cite{Barr:1981qv}, the gauge bosons $X$, in one case, can transform quarks to anti-quarks
assigned to the \textbf{10} representation; in the other case, they couple to
quarks and leptons in the \textbf{5} representation.
As a result, they can be resonantly produced at hadron colliders and decay into a top and a neutrino.
Therefore, any discovery of such signal imply new physics, and may help us to explore
the fundamental questions mentioned above.

In this work, we propose the general model-independent renormalizable effective Lagrangian
with $\rm SU(3)_c\times SU(2)_L\times U(1)_Y$  gauge symmetry
\begin{equation}\label{eq:L}
    \mathcal{L}=\la_S^{ij} \ep^{\al \be \ga}\phi_{\al}\bar{d}_{i \be R}^c d_{j \ga R}+
    a_S^i\phi_{\al}\bar{u}^{\al}_{iR}\chi
    +\la_V^{ij} \ep^{\al \be \ga} X_{\mu,\al}\bar{d}_{i \be R}^c \ga^{\mu} d_{j \ga R}+
    a_V^i X_{\mu,\al}\bar{u}^{\al}_{iR}\ga^{\mu}\chi+h.c.,
\end{equation}
where there is a summation over the generation indices $i,j=1,2,3,$ and
$\rm SU(3)_c$ gauge indices $\al,\be,\ga=1,2,3$. The superscript $c$
denotes charge conjugation. The Dirac field $\chi$ is a singlet under
the SM gauge group and manifest itself as missing energy at colliders. The
scalar and vector fields $\phi$ and $X_{\mu}$ are color triplet resonances that can
appear in some models, which obtain their masses at high energy scales.
This Lagrangian could further be generalized, such
as shown in Ref. \cite{Andrea:2011ws}, although it may not be  gauge invariant any more.
The free parameters in Eq. (\ref{eq:L}) are masses of the  resonances
and missing particle, i.e., $m_{\phi}, m_{X}$ and $m_{\chi}$, and couplings
$\la_{S,V}^{ij}$ and $a_{S,V}^i$, which should be constrained by current precise
data, and will be investigated carefully in this paper.
Here, we only consider the case of scalar resonance
field $\phi$, and the case of vector resonance field $X_{\mu}$ will be studied elsewhere.

The scenario of monotop production has been explored in Ref. \cite{Andrea:2011ws}, where
they only consider the mode of top hadronic decay. In the case of resonant monotop production,
they assume the branching fraction of $\phi\to t\chi$ equal to one and neglect the decay channel
of $\phi\to \bar{d} \bar{s}$, which would lead to an overestimation of the
signal. But we will take into account all decay channels of the resonance,
which turns out to be very important for estimating the sensitivity to detect the signal
at the LHC. Moreover, we also discuss the mode of semileptonic decay of top quark besides hadronic decay.
Although the cross section of the backgrounds for semileptonic decay mode are very large,
the discovery of the signal in this mode is still possible once appropriate cuts are imposed.

This paper is organized as follows. In Sec.~\ref{sec:exp},
we consider the constraints on the free parameters from $Z$ hadronic decay branching ratio,
$K^0-\overline{K^0}$ mixing and dijet experiments at the LHC.
In Sec.~\ref{sec:signal}, we investigate the signal and backgrounds of monotop production
in detail and then analyze the discovery potential at the early LHC.
A conclusion is given in Sec.~\ref{sec:conclusion}.

\section{Experiment Constraints}
\label{sec:exp}

The experiments have set constraints on the stop production and decay,
the signal of which is similar to the monotop,
in R-parity violating SUSY so far. For example, the H1 \cite{Aktas:2004ij} and ZEUS \cite{:2006je}
collaborations at HERA have analyzed the process of stop resonantly produced by electron-quark fusion and
%via the $\la^{\prime}_{131}$ coupling
followed either by a direct R-parity-violating decay, or by the gauge boson decay.
The process of stop pair production and decaying into dielectron plus dijet at the Tevatron
%,induced by the term with $\la^{\prime}_{13j}$,
is also discussed \cite{Chakrabarti:2003wr}.
However, these results can not be converted to constraints on the  parameters in our case.
Here the relevant experiments, we are concerned with, are $Z$ hadronic decay branching ratio,
$K^0-\overline{K^0}$ mixing and dijet production at the LHC.

\subsection{$Z$ hadronic decay branching ratio}

\begin{figure}
  % Requires \usepackage{graphicx}
  \includegraphics[width=0.8\linewidth]{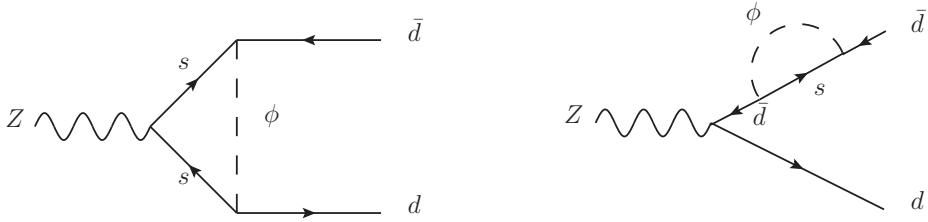}\\
  \caption{Feynman diagrams for hadronic $Z$ boson decay induced by the field $\phi$.}
  \label{eps:zdecay}
\end{figure}

The effective Lagrangian in Eq. (\ref{eq:L}) may contribute to the branching fraction of
$Z$ boson hadronic decay
as shown in Fig. \ref{eps:zdecay}.
 From the precise measurement of branching fraction of $Z$ boson hadronic decay,
 the relevant bands on R-parity violating SUSY parameters have been investigated
 in Ref. \cite{Bhattacharyya:1995bw}.
Since the quarks in the effective Lagrangian are
right-handed, the couplings of  right-handed quarks with $Z$ boson are modified,
and thus affect the branching fraction of $Z$ boson hadronic decay.

The tree-level amplitude of $Z$ boson decaying into a pair of quarks in the SM can be parameterized as
\begin{equation}\label{eqs:LOZdecay}
    \mathcal{M}_{\mu}=g_Z \bar{q}(p_1)\ga^{\mu}(a_L^q P_L + a_R^q P_R)q(p_2),
\end{equation}
where
\begin{eqnarray}
% \nonumber to remove numbering (before each equation)
  g_Z &=& \frac{e}{s_W c_W}, \nn\\
  a_L^q &=& t_3^q-Q_q s_W^2, \nn\\
  a_R^q &=& -Q_q s_W^2.
\end{eqnarray}

After calculating the Feynman diagrams in Fig. \ref{eps:zdecay}, we find that
the coefficient $a_R^q$ is adjusted by multiplying a factor
\begin{equation}
    1+\De_f=1+\frac{\la_f^2}{8\pi^2}g(a),
\end{equation}
where $a=M_Z^2/m_{\phi}^2$ and $f=1,2,3$ correspond to $Z$ boson decaying
into $d\bar d$, $s\bar s$ and $b\bar b$, respectively.
And $\la_f$ are defined as
\begin{eqnarray}
% \nonumber to remove numbering (before each equation)
  \la_1^2 &=& 4[(\la_S^{12})^2+(\la_S^{13})^2], \\
  \la_2^2 &=& 4[(\la_S^{12})^2+(\la_S^{23})^2], \\
  \la_3^2 &=& 4[(\la_S^{13})^2+(\la_S^{23})^2],
\end{eqnarray}
where we have used the fact that $\la_S^{ij}=-\la_S^{ji}$ due to the
antisymmetry of the $\ep^{\al \be \ga}$ couplings in Eq. (\ref{eq:L}).
The explicit form of function $g(a)$ is
\begin{eqnarray}\label{eq:ga}
% \nonumber to remove numbering (before each equation)
  g(a) &=&\frac{(a-4) a-2 \log (a) ((a-2) a+2 \log (a+1))-4
   \text{Li}_2(-a)}{4 a^2}.
\end{eqnarray}
The ultraviolet poles of the triangle and self-energy diagrams have canceled each other, and we
obtain a finite result. In this calculation, all the masses of quarks are neglected.
Eq. (\ref{eq:ga}) seems divergent if $a$ vanishes due to the denominator $a^2$. But actually we expand
this result around $a=0$, and get the asymptotic form
\begin{equation}
   g(a)= \left(\frac{1}{9}-\frac{\log (a)}{3}\right)
   a+\left(\frac{\log (a)}{4}-\frac{1}{16}\right)
   a^2+\left(\frac{1}{25}-\frac{\log (a)}{5}\right)
   a^3+O\left(a^4\right),
\end{equation}
which vanishes obviously when taking the limit $a\to 0$. This feature guarantees the decouple of the heavy
particle $\phi$ in the large $m_{\phi}$ limit.

There are two observables which can be affected by the change of coefficient $a_R^q$.
One is $R_l\equiv \Ga_h/\Ga_l$, where $\Ga_{h,l}$ are the widths of $Z$ boson decaying into
 hadrons and leptons, respectively. The correction to $R_l$ is
\begin{eqnarray}
% \nonumber to remove numbering (before each equation)
  \de R_l &=& \frac{\Ga_h-\Ga_h^{SM}}{\Ga_l^{SM}} \nn\\
   &=& \frac{2(\De_1\Ga_{dR}^{SM}+\De_2\Ga_{sR}^{SM}+\De_3\Ga_{bR}^{SM})}{\Ga_l^{SM}} ,
\end{eqnarray}
where $\Ga_{qR}^{SM},q=d,s,b$ denote the widths of $Z$ boson decaying into only right-handed $q$ quarks in the SM.
The other is $R_b\equiv \Ga_b/\Ga_h$, where
$\Ga_b$ is the width into $b\bar b$.
Explicitly, we can write  $R_b$ as
\begin{eqnarray}
% \nonumber to remove numbering (before each equation)
  R_b &=& \frac{\Ga_b}{\Ga_h} \nn \\
   &=& \frac{1+2\De_3\frac{\Ga_{bR}^{SM}}{\Ga_b^{SM}}}
   {1+2\De_1\frac{\Ga_{dR}^{SM}}{\Ga_h^{SM}}+2\De_2\frac{\Ga_{sR}^{SM}}{\Ga_h^{SM}}+2\De_3\frac{\Ga_{bR}^{SM}}{\Ga_h^{SM}}}
   \frac{\Ga_b^{SM}}{\Ga_h^{SM}}.
\end{eqnarray}
Thus, the correction to $R_b$ is given by
\begin{eqnarray}
% \nonumber to remove numbering (before each equation)
  \de R_b &\approx& 2\left[\De_3 \frac{\Ga_{bR}^{SM}}{\Ga_b^{SM}}\left(1-\frac{\Ga_b^{SM}}{\Ga_h^{SM}}\right)
  -\De_1 \frac{\Ga_{dR}^{SM}}{\Ga_d^{SM}}\frac{\Ga_d^{SM}}{\Ga_h^{SM}}
  -\De_2 \frac{\Ga_{sR}^{SM}}{\Ga_s^{SM}}\frac{\Ga_s^{SM}}{\Ga_h^{SM}}
   \right]R_b^{SM}.
\end{eqnarray}
The experiments give $R_e=20.804\pm 0.050$,
$R_{\mu}=20.785\pm 0.033$, $R_{\tau}=20.764\pm 0.045$ and $R_{b}=0.2163\pm 0.0007$, respectively,
while the SM predictions are
$R_e^{SM}=R_{\mu}^{SM}=20.735$, $R_{\tau}^{SM}=20.780$ and $R_{b}^{SM}=0.2158$ \cite{Nakamura:2010zzi}.
The requirement that the corrected $R_{e,\mu,\tau,b}$ are in the $1\sigma$ range around the experimental central values imposes
constraints as follows,
\begin{eqnarray}\label{eqs:constraint-Z1}
    (\la_1^2+\la_2^2+\la_3^2)\frac{g(a)}{4\pi^2}< 0.829, 0.578, 0.202 {~~~\rm for} ~~R_{e}, R_{\mu}, R_{\tau}, {\rm respectively},
\end{eqnarray}
and
\begin{eqnarray}
    -0.0289<[0.78\la_3^2-0.22(\la_1^2+\la_2^2)]\frac{g(a)}{4\pi^2} < 0.173 {~~~\rm for} ~~R_{b}.
\end{eqnarray}

\begin{figure}
  % Requires \usepackage{graphicx}
  \includegraphics[width=0.5\linewidth]{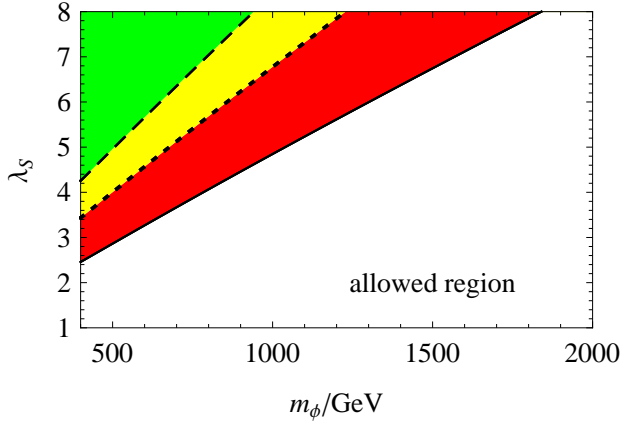}\\
  \caption{The allowed region
  by $Z$ boson hadronic decay branching fraction as a function of $m_{\phi}$.
  The solid and dashed lines are the upper limits given by $R_{\tau}$
for the cases $\la_S^{12}=\la_S^{13}=\la_S^{23}=\la_S$ and $\la_S^{12}=\la_S,\la_S^{13}=\la_S^{23}=0$, respectively.
The dotted line is the upper limit given by $R_b$ for $\la_S^{12}=\la_S,\la_S^{13}=\la_S^{23}=0$.}
  \label{eps:para}
\end{figure}

We show the
allowed region by $R_{\tau}$ and $R_b$ for $\la_S$ as a function of $m_{\phi}$ in Fig. \ref{eps:para}.
The solid and dashed lines are the upper limits given by $R_{\tau}$
for the cases $\la_S^{12}=\la_S^{13}=\la_S^{23}=\la_S$ and $\la_S^{12}=\la_S,\la_S^{13}=\la_S^{23}=0$, respectively.
The dotted line is the upper limit given by $R_b$ for $\la_S^{12}=\la_S,\la_S^{13}=\la_S^{23}=0$.
From Fig. \ref{eps:para} we can see that this constraint on the parameter is not very stringent.
This is due to the fact that only right-handed couplings are corrected, and
the widths of $Z$ boson decaying into right-handed quarks are much less than into left-handed quarks.

\subsection{$K^0-\overline{K^0}$ mixing}

\begin{figure}
  % Requires \usepackage{graphicx}
  \includegraphics[width=0.5\linewidth]{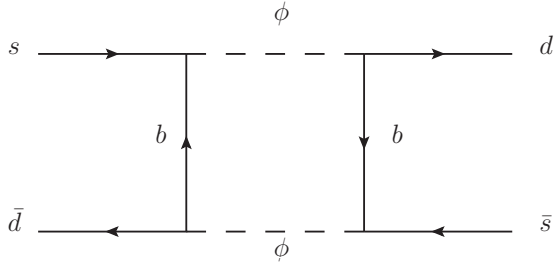}\\
  \caption{Representative Feynman diagram for $K^0-\overline{K^0}$ mixing.}
  \label{eps:KKmixing}
\end{figure}

Now we consider the constraint from $K^0-\overline{K^0}$ mixing.
The typical Feynman diagram for $K^0-\overline{K^0}$ mixing is shown in Fig. \ref{eps:KKmixing}.
After straightforward calculations, we can obtain
\begin{equation}
    \mathcal{H}_{eff}^{\De S=2}=C Q,
\end{equation}
where $Q$ is the operator $\bar{d}_R^\al\ga^{\mu}s_R^{\al}\bar{d}_R^\be\ga^{\mu}s_R^{\be}$, and
$C$ is its Wilson coefficient,
\begin{equation}
    C=\frac{(\la_S^{13}-\la_S^{31})^2(\la_S^{23}-\la_S^{32})^2}{32\pi^2}
    \left[\frac{m_{\phi}^4-m_b^4-2m_b^2 m_{\phi}^2 \ln\frac{m_{\phi}^2}{m_b^2}}{(m_{\phi}^2-m_b^2)^3}\right]b(\mu),
\end{equation}
where
\begin{equation}
    b(\mu)=(\al_s(\mu))^{-2/9}\left(1+\frac{307}{162}\frac{\al_s(\mu)}{4\pi}\right)
\end{equation}
contains the renormalization scale dependence \cite{Herrlich:1995hh}.
We have compared this result with that in Refs. \cite{deCarlos:1996yh,Slavich:2000xm}
and find our result is consistent with their results.
Then, the $K_L-K_S$ mass difference $\De m_K$ is given by \cite{Ciuchini:1998ix}
\begin{equation}
    \De m_K=2 {\rm Re} \langle K^0|\mathcal{H}_{eff}^{\De S=2}|\overline{K^0}\rangle
    =2 C{\rm Re} \langle K^0| Q|\overline{K^0}\rangle.
\end{equation}
The matrix element $\langle K^0|Q|\overline{K^0}\rangle$ can be parameterized as
\begin{equation}
    \langle K^0|Q|\overline{K^0}\rangle=\frac{1}{3}m_K f_K^2 B_K(\mu)
\end{equation}
where $m_K$ is the mass of $K^0$ (497.6~MeV), $f_K$ is kaon decay constant (160~MeV),
and $B_K(\mu)$ is related to the renormalization group invariant parameter $\hat{B}_K$ by
\begin{equation}
    \hat{B}_K=B_K(\mu)b(\mu).
\end{equation}
In our numerical analysis we will use the following result \cite{Buras:1998raa}:
\begin{equation}
    \hat{B}_K=0.75\pm0.15.
\end{equation}
On the other hand, the SM contribution to $\De m_K$ is
\begin{equation}
    \De m^{SM}_K=\frac{G_F^2}{6\pi^2}f_K^2 \hat{B}_K m_K M_W^2 {\rm Re}
    [\la_c^{\ast2}\eta_1 S_0(x_c)+\la_t^{\ast 2}\eta_2 S_0(x_t)+2\la_c^{\ast}\la_t^{\ast}\eta_3 S_0(x_c,x_t)]
\end{equation}
where $\la_i=V_{is}^{\ast}V_{is}$, and $V_{ij}$ are the CKM matrix elements.
The functions $S_0$ are given by
\begin{eqnarray}
% \nonumber to remove numbering (before each equation)
  S_0(x_t) &=& \frac{4x_t-11x_t^2+x_t^3}{4(1-x_t)^2}-\frac{3x_t^3\ln x_t}{2(1-x_t)^3},\nn \\
  S_0(x_c) &=& x_c,\nn \\
  S_0(x_c,x_t) &=& x_c\left[\ln \frac{x_t}{x_c} -\frac{3x_t}{4(1-x_t)}-\frac{3x_t^2\ln x_t}{4(1-x_t)^2}\right]
\end{eqnarray}
with $x_i=m_i^2/M_W^2$.
The next-to-leading values of $\eta_i$ are given as follows \cite{Buras:1990fn,Herrlich:1993yv,Urban:1997gw}:
\begin{equation}
    \eta_1=1.38\pm 0.20, \quad \eta_2=0.57\pm 0.01, \quad \eta_3=0.47\pm 0.04.
\end{equation}

\begin{figure}
  % Requires \usepackage{graphicx}
  \includegraphics[width=0.5\linewidth]{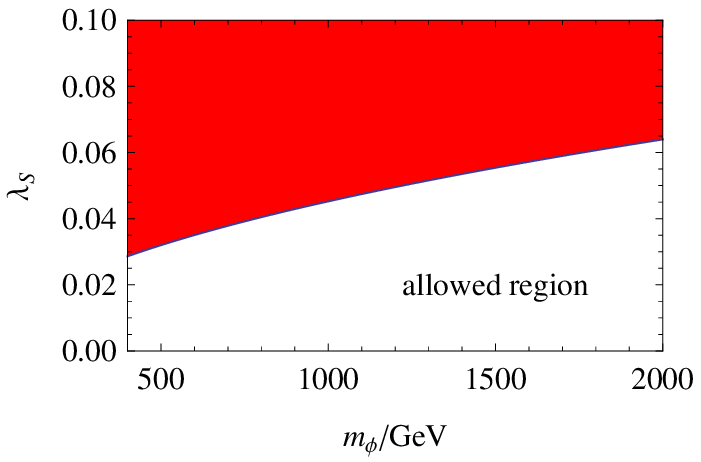}\\
  \caption{The allowed region of $\la_S$ ($=\la_S^{13}=\la_S^{23}$) by
  $K^0-\overline{K^0}$ mixing as a function of $m_{\phi}$.}
  \label{eps:paraKKmixing}
\end{figure}

We require that the contribution to $\De m_K$, including the SM and new physics result,
is not larger than the experimental value
$\De m_K^{\rm exp}=(3.483\pm0.006)\times 10^{-15}~\rm GeV$ \cite{Nakamura:2010zzi}
within $1\sigma$ level,
assuming the CPT conservation.
In Fig. \ref{eps:paraKKmixing}, we show the
allowed region for $\la_S$ as a function of $m_{\phi}$
for $\la_S^{13}=\la_S^{23}=\la_S$.
From Fig. \ref{eps:paraKKmixing} we find that the constraint on $\la_S$ is very stringent,
generally less than 0.06. Furthermore, these couplings involves the third generation quarks,
the parton distribution functions (PDFs) of which are small compared with those of the first two
generations. Therefore, we choose $\la_S^{13}=\la_S^{23}=0$,
for simplicity, in the following discussion.

\subsection{Dijet production at the LHC}
\begin{figure}
  % Requires \usepackage{graphicx}
  \includegraphics[width=0.5\linewidth]{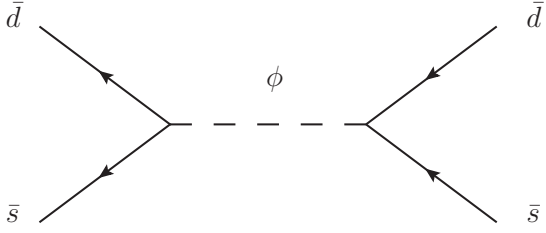}\\
  \caption{Feynman diagram for the dijet production.}
  \label{eps:dijet}
\end{figure}

The dijet experiments at the LHC have set upper limits on the product of cross section ($\sigma_{jj}$)
and signal acceptance ($\mathcal{A}$)
for resonance productions \cite{Khachatryan:2010jd,:2010bc,Aad:2011aj,Chatrchyan:2011ns,Aad:2011fq},
such as excited quarks, axigluons, Randall-Sundrum gravitons,
diquarks and string resonances.
We can use these data to constrain the parameters in the effective Lagrange in Eq. (\ref{eq:L}).
The relevant Feynman diagram for the dijet production is shown in Fig. \ref{eps:dijet}.

\begin{figure}
  % Requires \usepackage{graphicx}
  \includegraphics[width=0.5\linewidth]{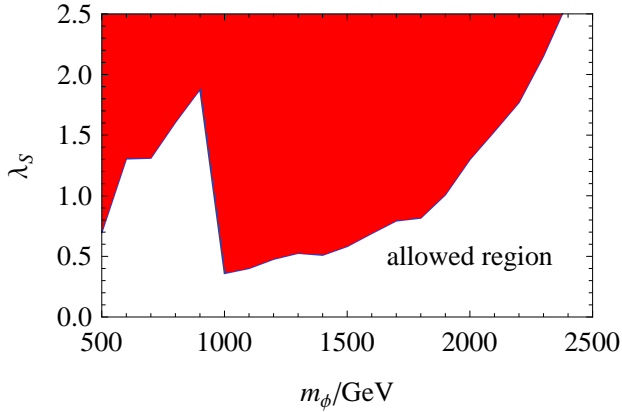}\\
  \caption{The allowed region of $\la_S(=\la_S^{12}=a_S^3)$
  by dijet experiments at the LHC as a function of $m_{\phi}$.}
  \label{eps:para2j}
\end{figure}

The cross section of the resonance $\phi$ production and decaying into dijet is highly sensitive
to the width of $\phi$ decay, which is given by
\begin{equation}
    \Ga_{\phi}=\Ga_{\phi\to \bar d \bar s}+\Ga_{\phi\to u_i \bar \chi},
\end{equation}
 with
\begin{eqnarray}\label{eq:phiwidth}
    \Ga_{\phi\to \bar d \bar s}&=&\frac{(\la_S^{12})^2}{2\pi}m_{\phi}, \nn\\
    \Ga_{\phi\to u_i \bar \chi}&=&\frac{|a_S^i|^2}{16\pi m_{\phi}^3}(m_{\phi}^2-m_{u_i}^2-m_{\chi}^2)
    \la^{1/2}(m_{\phi}^2,m_{u_i}^2,m_{\chi}^2),
\end{eqnarray}
where $\la(a,b,c)=a^2+b^2+c^2-2ab-2bc-2ca$.
We will take into account the effect of these widths in our numerical calculation below.
To calculate the cross section,
we use MadGraph5v1.3.3 \cite{Alwall:2011uj} with the effective Lagrangian implemented in by
FeynRules \cite{Christensen:2008py}.
We vary the mass of $\phi$ from 500 GeV to 2500 GeV with a step of 100 GeV. For each mass, we
calculate the decay width of $\phi$, assuming $\la_S^{12}=a_S^3=0.2, a_S^1=a_S^2=0, m_{\chi}=50{~\rm GeV}$.
Then we change the corresponding parameters in MadGraph and calculate the cross sections
of the dijet production.
We choose the kinematical cuts as following \cite{Khachatryan:2010jd,Chatrchyan:2011ns}:
\begin{eqnarray}\label{eq:cuts-dijet}
% \nonumber to remove numbering (before each equation)
   &&|\eta_j| < 2.5, \quad     |\eta_{j_1}-\eta_{j_2}| < 1.3.
\end{eqnarray}
The cross sections of the dijet signal before and after the cuts are listed in Table. \ref{tab:dijetCrossSection}.
\begin{table}
  \centering
  \begin{tabular}{lccccccc}
  \hline\hline
  % after \\: \hline or \cline{col1-col2} \cline{col3-col4} ...
  $m_{\phi}$ (GeV) & 500 & 600 & 700 & 800 & 900 & 1000 & 1100\\
  \hline\hline
  $\sigma_0$ (pb)&28.2&12.6&6.11&3.17&1.73&9.80$\times 10^{-2}$&5.71$\times 10^{-2}$\\
  $\sigma_f$ (pb)&16.2&7.13&3.52&1.84&9.98$\times 10^{-2}$&5.66$\times 10^{-2}$&3.22$\times 10^{-2}$ \\
  \hline\hline
  % after \\: \hline or \cline{col1-col2} \cline{col3-col4} ...
  $m_{\phi}$ (GeV)  & 1200 & 1300 & 1400 & 1500 & 1600 &1700 &1800\\
  \hline\hline
  $\sigma_0$ (pb)&3.40$\times 10^{-2}$&2.07$\times 10^{-2}$&1.28$\times 10^{-2}$&7.95$\times 10^{-2}$
  &5.00$\times 10^{-2}$&3.17$\times 10^{-2}$&2.03$\times 10^{-2}$ \\
  $\sigma_f$ (pb)&1.94$\times 10^{-2}$&1.18$\times 10^{-2}$&7.27$\times 10^{-2}$&4.55$\times 10^{-2}$
  &2.82$\times 10^{-2}$&1.79$\times 10^{-2}$&1.16$\times 10^{-2}$ \\
  \hline\hline
  % after \\: \hline or \cline{col1-col2} \cline{col3-col4} ...
  $m_{\phi}$ (GeV)  & 1900 & 2000 & 2100 & 2200 & 2300 &2400 &2500\\
  \hline\hline
  $\sigma_0$ (pb)&1.30$\times 10^{-2}$&8.35$\times 10^{-3}$&5.38$\times 10^{-3}$&3.47$\times 10^{-3}$
  &2.23$\times 10^{-3}$&1.44$\times 10^{-3}$&9.27$\times 10^{-4}$ \\
  $\sigma_f$ (pb)&7.42$\times 10^{-3}$&4.79$\times 10^{-3}$&3.07$\times 10^{-3}$&1.99$\times 10^{-3}$
  &1.28$\times 10^{-3}$&8.35$\times 10^{-4}$&5.25$\times 10^{-4}$ \\
  \hline
  \end{tabular}
  \caption{The cross sections of dijet production induced by the resonance of $\phi$
  before ($\sigma_0$) and after ($\sigma_f$) the cuts given in Eq. (\ref{eq:cuts-dijet}),
  assuming $\la_S^{12}=a_S^3=0.2, a_S^1=a_S^2=0$.}
  \label{tab:dijetCrossSection}
\end{table}
 Fig. \ref{eps:para2j} shows the allowed region of $\la_S(=\la_S^{12}=a_S^3)$ as a function of $m_{\phi}$, where
we choose the acceptance $\mathcal{A}=0.6$ as in Ref. \cite{Chatrchyan:2011ns}.
It is required that $\sigma_{jj}\cdot \mathcal{A}$ is not larger than
the observed $95\%$ C.L. upper limit in the dijet experiment \cite{Khachatryan:2010jd,Chatrchyan:2011ns}.
The bump of the curve in the region from 500 GeV to 1000 GeV for $m_{\phi}$ is due to the fact
that we compare with data in this region and the other regions corresponding to
integrated luminosities of 2.9 ${\rm pb^{-1}}$ and 1 ${\rm fb^{-1}}$, respectively,
 collected by the CMS experiment at the LHC.

\section{Signal and Background}
\label{sec:signal}
\begin{figure}
  % Requires \usepackage{graphicx}
  \includegraphics[width=0.8\linewidth]{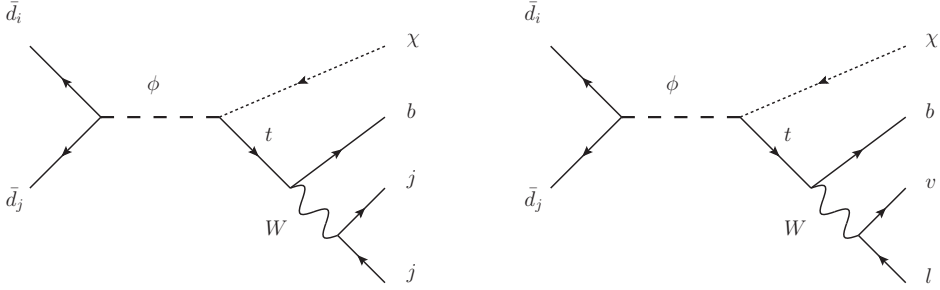}\\
  \caption{Feynman diagrams for the monotop production.}
  \label{eps:Feynman}
\end{figure}

The signals of the monotop production are
\begin{eqnarray}
 %\nonumber to remove numbering (before each equation)
  pp\to t+\Slash{E}_T\to bW + \Slash{E}_T\to bjj + \Slash{E}_T \quad {\rm and}\quad  bl + \Slash{E}_T,
\end{eqnarray}
which are shown in Fig. \ref{eps:Feynman}.
The symbol $b$ and $j$ denote a $b$-tagged jet and light quark or gluon jet, respectively,
and $l$ refers to the first two generation charged leptons, i.e., $e$ and $\mu$.
We define the process with top hadronic decay as hadronic mode,
while the one with top semileptonic decay as semileptonic mode.
The hadronic mode suffers from fewer backgrounds in the SM than the
semileptonic mode because of
the smaller phase space due to more particles in the final states.
This mode has been studied in Ref. \cite{Andrea:2011ws} where
they assume the branching fraction $R(\phi\to t\bar \chi)$ equal to one.
However, this assumption is over optimistic.
From Eq. (\ref{eq:phiwidth}) we get the branching fraction $R(\phi\to t\bar \chi)$,
\begin{eqnarray}
% \nonumber to remove numbering (before each equation)
  R(\phi \to t\bar\chi) &=& \frac{\Ga_{\phi \to t\bar \chi}}{\Ga_{\phi \to t\bar \chi}+\Ga_{\phi \to\bar d \bar s}} = \frac{1}{1+z},
\end{eqnarray}
with
\begin{eqnarray}
% \nonumber to remove numbering (before each equation)
  z &=& \frac{8(\la_S^{12})^2}{|a_S^3|^2} \frac{m_{\phi}^4}{(m_{\phi}^2-m_t^2-m_{\chi}^2)\la^{1/2}(m_{\phi}^2,m_{t}^2,m_{\chi}^2)}.
\end{eqnarray}
Here we assume that the decay widths $\Ga_{\phi \to u\bar \chi}=\Ga_{\phi \to c\bar \chi}=0$.
In the case of $\la_S^{12}=a_S^3=0.2,m_t=173.1{~\rm GeV}, m_{\phi}=500{~\rm GeV} {\rm ~and~} m_{\chi}=50{~\rm GeV}$,
we find $\Gamma_{\phi \to t \bar{\chi}}=0.300 {~\rm GeV}$,
$\Gamma_{\phi \to \bar{d} \bar{s}}=3.183 {~\rm GeV}$, and the branching fraction of $\phi\to t\bar{\chi}$ is just about 0.1.
So, in this work, we take into account the effect of both $\phi$ decay channels
and below we will discuss further the hadronic and leptonic modes in detail.

Before discussing the signal and backgrounds in detail, we first give some comments
on the parameter $m_{\chi}$. In the SUSY model,
without the assumption of gaugino mass unification, there is no general mass limit
from $e^+e^-$ colliders for the lightest neutralino~\cite{Nakamura:2010zzi}.
The indirect constraints from $(g-2)_{\mu}$, $b\to s\gamma$ and $B\to \mu^+\mu^-$
show that the lightest neutralino mass can be as low as about 6 GeV~\cite{Belanger:2003wb}.
In our case, we choose the default value of $m_{\chi}=50$ GeV and discuss the
effect on the discovery significance when varying $m_{\chi}$ in the range
$5-100$ GeV.  An estimate of the width can be made
by the Feynman diagrams shown in Fig. \ref{eps:Feynman},
where we can consider only $\chi$ as the initial-state particle, for example,
\begin{equation}
\chi(p_1) \to d(p_2) s(p_3) b(p_4) \nu (p_5)l^+(p_6).
\end{equation}
Then the width of $\chi$ is given by
\begin{equation}\label{eqs:width}
    \Gamma_{\chi} = \frac{1}{2m_{\chi}}\int |\overline{\mathcal{M}}|^2 d \Gamma_5,
\end{equation}
where $|\overline{\mathcal{M}}|^2$ is the matrix element squared for the decay process
which has taken into account the average and sum over the initial- and final-state spins and colors.
When the masses of all the final-state particles are neglected,
the five body phase space integration can be written as
\begin{eqnarray}
% \nonumber to remove numbering (before each equation)
  \int d \Gamma_5 &=& \frac{1}{32768\pi^7}\frac{1}{m_{\chi}^2}
  \int_0^{m_{\chi}^2}ds_{23}
  \int_0^{(m_{\chi}-\sqrt{s_{23}})^2}ds_{456}
  \int_0^{s_{456}}ds_{45} \nn  \\
  && \lambda^{1/2}(m_{\chi}^2,s_{23},s_{456})
  \left(1-\frac{s_{45}}{s_{456}}\right),
\end{eqnarray}
where $s_{ij}=(p_i+p_j)^2$ and $s_{ijk}=(p_i+p_j+p_k)^2$.
In the mass range of $\chi$ we are interested in,
the momenta of the decay products of the $W$ boson are so small compared with
the mass of the $W$ boson that we neglect them in the calculation of the matrix element.
Moreover, we assume that the lepton $l^+$ carries about one-fifth of the energy of $\chi$ on average.
In this case, the matrix element squared is simply given by
\begin{equation}
    |\overline{\mathcal{M}}|^2\approx\frac{96}{5}g_W^4(\lambda_S^{12})^2 (a_S^3)^2
    \frac{m_{\chi}^2}{m_{\phi}^4 m_t^2 M_W^4}s_{23}s_{45}.
\end{equation}
where $g_W$ is the coupling of the $W$ boson with left-handed fermions.
Then we perform the integration in Eq. (\ref{eqs:width}), and obtain
\begin{equation}\label{eqs:chiwidth}
    \Gamma_{\chi} \approx 1.82\times 10^{-19} {\rm GeV}\left(\frac{\lambda_S^{12}}{0.2}\right)^2 \left(\frac{a_S^3}{0.2}\right)^2
    \frac{(m_{\chi}/50 {\rm GeV})^{11}}{(m_{\phi}/500 {\rm GeV})^4 (m_t/173.1 {\rm GeV})^2 (M_W/80.4 {\rm GeV})^4}.
\end{equation}
\begin{figure}
  % Requires \usepackage{graphicx}
  \includegraphics[width=0.5\linewidth]{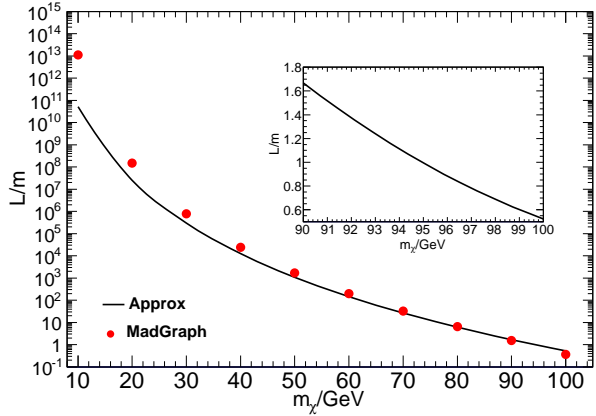}\\
  \caption{The distance travelled by the particle $\chi$  before its decay as a function of its mass.
  The solid line is obtained from Eq. (\ref{eqs:chiwidth}) while the dots denote the results of MadGraph.
  The relevant parameters are chosen as $\la_S^{12}=a_S^3=0.2$, $m_t=173.1{~\rm GeV}, M_W=80.4{~\rm GeV},{\rm ~and~} m_{\phi}=500{~\rm GeV}$.}
  \label{eps:length}
\end{figure}
The produced $\chi$ at hadron colliders, as a decay product of a massive particle,
usually has such a large energy that it moves nearly in the speed of light.
In Fig. \ref{eps:length}, we show the distance travelled by the particle $\chi$
before its decay as a function of its mass.
It can be seen that the distance strongly depends on the mass of  $\chi$
and decreases with increasing $m_{\chi}$.
The results of MadGraph are well approximated by those obtained from Eq. (\ref{eqs:chiwidth})
except for the low mass region since we have neglected the mass of final-state particle in Eq. (\ref{eqs:chiwidth}).
But this discrepancy between them in the low mass region is not important
because they are both much larger than the size of the detector at the LHC.
The ATLAS collaboration has searched for displaced vertices
arising from decays of new heavy particles and
found that the efficiency for detecting displaced vertices almost vanishes
for a distance between the primary and the displaced vertex larger than 0.35 m \cite{Aad:2011zb}.
Therefore, as shown in Fig. \ref{eps:length},  it is reasonable that  the particle $\chi$
with a mass less than 100 GeV is  considered as missing energy at the LHC.

\subsection{Hadronic mode}
For the hadronic mode, the main backgrounds arise from $pp\to jjjZ( \nu \bar{\nu})$,
with a jet misidentified as a $b$-jet, and  $pp\to b\bar{b}jZ(\nu \bar{\nu})$
with a $b$-jet not tagged.
The top pair and single top production processes
with hadronic top quark decay may also contribute to the backgrounds if some jets are not detected.
The signal and backgrounds are simulated by MadGraph5v1.3.3 \cite{Alwall:2011uj}
and ALPGEN \cite{Mangano:2002ea} interfaced with PYTHIA \cite{Sjostrand:2007gs,Sjostrand:2006za}
to perform the parton shower and hadronization.
In this mode, the momentum of three jets, and therefore momentum of the $W$ boson and top quark,
can be reconstructed, which leads to efficient event selection using invariant mass cut.
In the following numerical calculation, the default relevant parameters are chosen as
$\la_S^{12}=a_S^3=0.2, \la_S^{13}=\la_S^{23}=0, a_S^1=a_S^2=0,
m_t=173.1{~\rm GeV}, m_{\phi}=500{~\rm GeV} {\rm ~and~} m_{\chi}=50{~\rm GeV}$, and
CTEQ6L1 PDF is used.
The renormalization and factorization scales are set at $m_{\phi}$.
We use the following basic selection cuts
\begin{eqnarray}
% \nonumber to remove numbering (before each equation)
  p_T^{b,j} > 30{~\rm GeV}, \quad |\eta^{b,j}| < 2.4,\quad  {\De R}_{bb,bj,jj} > 0.5.
 \end{eqnarray}\label{eqs:cuts}
Moreover, we choose a $b$-tagging efficiency of 50$\%$
while the misidentification rates for other jets are, 8$\%$ for charm quark, 0.2$\%$ for gluon
and other light quarks \cite{Aad:2009wy}.

\begin{figure}
  % Requires \usepackage{graphicx}
  \includegraphics[width=0.5\linewidth]{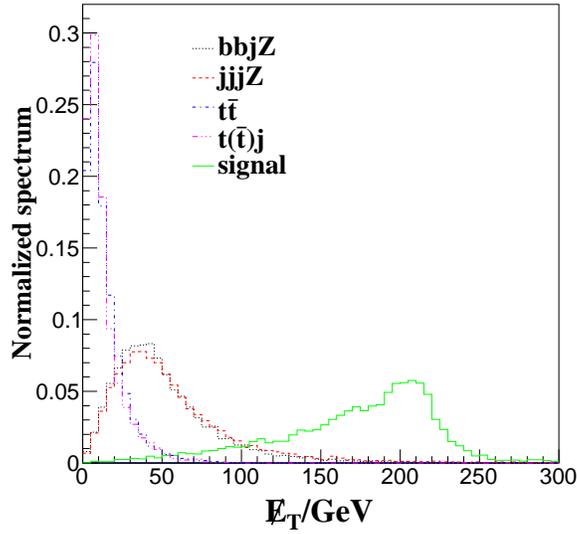}\\
  \caption{The normalized spectrum of missing transverse energy
  in the hadronic mode at the LHC ($\sqrt{s}=$ 7 TeV). }
  \label{eps:missing_pt}
\end{figure}

To determine the missing transverse energy cut, we show the normalized spectrum
of the missing transverse energy for the signal and backgrounds in Fig. \ref{eps:missing_pt}.
The backgrounds concentrate in the region $\Slash{E}_{T}<100 {~\rm GeV}$ because the
missing transverse energy of the background comes from either an invisible decayed $Z$ boson
or non-detected jets, which are produced mainly via t-channel.
In contrast, the missing transverse energy of the signal results from the decay of a heavy resonance
so that it can be large. Therefore we choose the missing transverse energy cut
\begin{equation}
\Slash{E}_{T}>100 {~\rm GeV}.
\end{equation}

Meanwhile, the shape of the signal is similar to the distribution
$\Slash{E}_T/\sqrt{{\Slash{E}_T^{max}}^2-\Slash{E}_T^2}$
with an edge at $\Slash{E}_T^{max}=\la^{1/2}(m_{\phi}^2,m_t^2,m_{\chi}^2)/2 m_{\phi}$.
This feature may help to specify the masses of the resonance and the missing particle.

\begin{figure}
  % Requires \usepackage{graphicx}
  %\includegraphics[width=0.45\linewidth]{mw.eps}
  \includegraphics[width=0.5\linewidth]{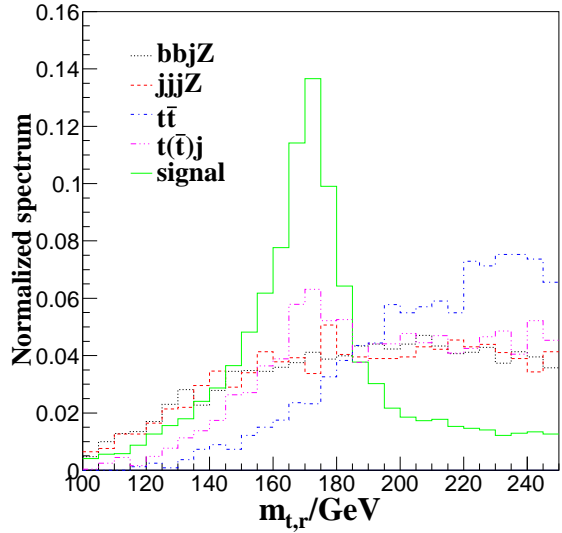}\\
  \caption{The reconstructed  top quark mass distribution for the signal and backgrounds processes.}
  \label{fig:mt}
\end{figure}

In Fig. \ref{fig:mt} we show the reconstructed top quark mass distribution
for the signal and backgrounds processes using the three leading jets.
It can be seen that there is a peak around $175$ GeV for
the signal while the distributions of backgrounds grow up with the increase of
reconstructed top quark mass,
and thus we impose the invariant mass cut in the final states as following,
\begin{eqnarray}
% \nonumber to remove numbering (before each equation)
    120~{\rm GeV} <  m_{t,r}  < 200~{\rm GeV}.
\end{eqnarray}

The cross sections of the signal and backgrounds   after various cuts at the LHC  ($\sqrt{s}=$ 7 TeV)
 are listed in Table~\ref{tab:hadronic}.
It can be seen that the backgrounds decrease dramatically when the invariant mass cuts are imposed,
and the cross section of $b\bar{b}jZ(\nu \bar{\nu})$ is not smaller than that of $ jjjZ(\nu \bar{\nu})$ after
all cuts imposed so that it can not be neglected.
The $t\bar{t}$ and $t(\bar{t})j$ processes are mainly suppressed by the missing transverse energy cut,
which can be seen from Fig. \ref{eps:missing_pt}.

\begin{figure}
  % Requires \usepackage{graphicx}
  \includegraphics[width=0.5\linewidth]{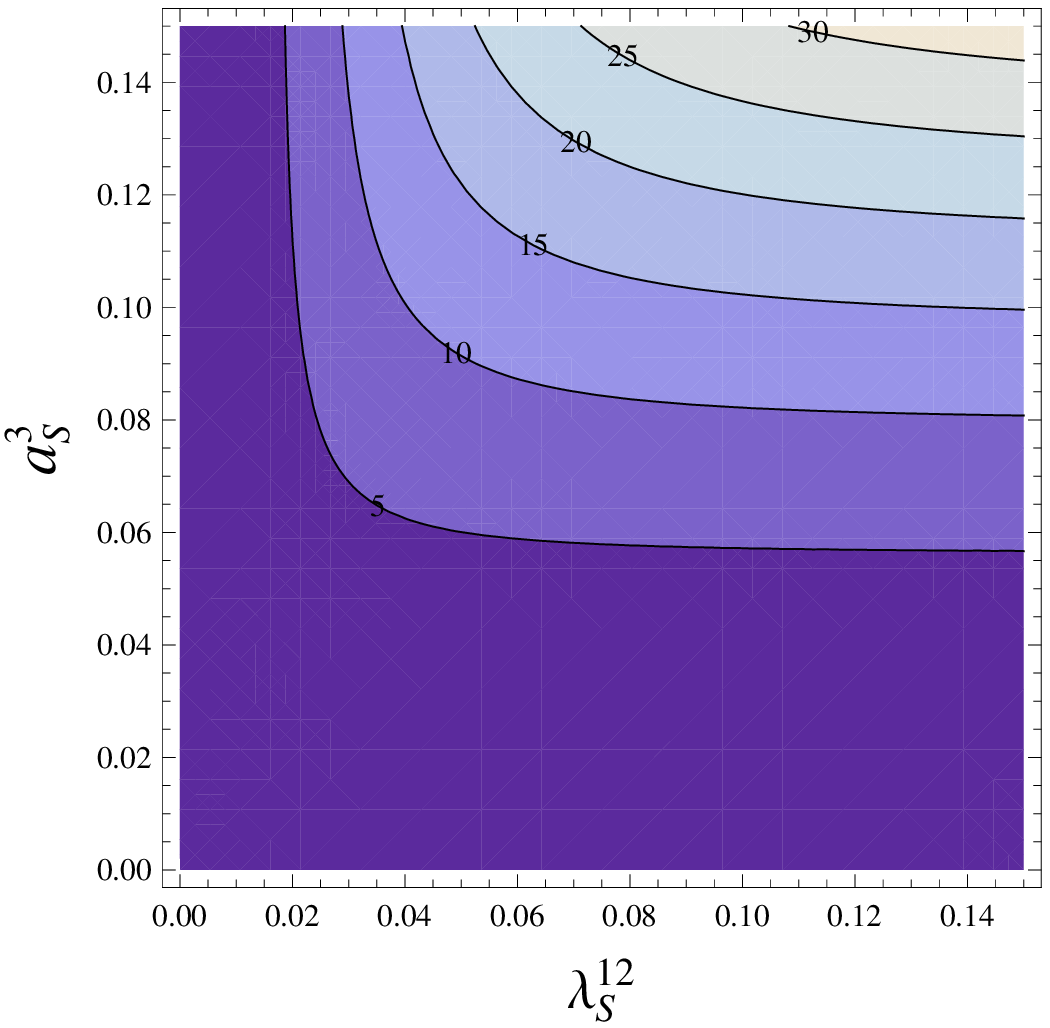}\\
  \caption{The significance  in the hadronic mode at the LHC ($\sqrt{s}=$ 7 TeV) with an integrated luminosity
of 1 $\rm fb^{-1}$ versus the parameters $\la_S^{12}$ and $a_S^3$,
assuming $m_{\phi}=500{~\rm GeV} {\rm ~and~} m_{\chi}=50{~\rm GeV}$. }
  \label{eps:signif}
\end{figure}

\begin{figure}
  % Requires \usepackage{graphicx}
  \includegraphics[width=0.5\linewidth]{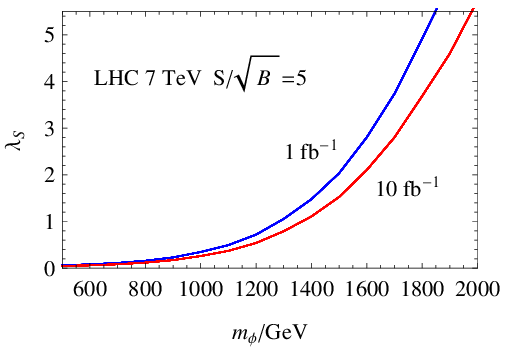}\\
  \caption{The $5\sigma$ discovery limits of $m_{\phi}$ and $\la_S(=\la_S^{12}=a_S^3)$
  in the hadronic mode at the LHC ($\sqrt{s}=$ 7 TeV).
  Either band consists of twenty solid lines from the bottom up
  corresponding to the value of $m_{\chi}$ varying from 5 GeV to 100 GeV with a step of 5 GeV. }
  \label{eps:sighrd}
\end{figure}

To investigate the discovery potential of monotops in the hadronic mode at the LHC ($\sqrt{s}=$ 7 TeV)
with an integrated luminosity of 1 $\rm fb^{-1}$, in Fig. \ref{eps:signif} we present the contour
curves of significance $\mathcal{S}=S/\sqrt{B}$ versus the parameters $\la_S^{12}$ and $a_S^3$,
where $S$ and $B$ are respectively the expected numbers of the signal and backgrounds events.
And in Fig. \ref{eps:sighrd} we present the $5\sigma$ ($\mathcal{S}=5$) discovery limits of
$m_{\phi}$, $m_{\chi}$ and $\la_S^{12}=a_S^3=\la_S$.
From Fig. \ref{eps:signif} we can see that for a $5\sigma$ discovery,
the sensitivity to $\la_S^{12}$ and $a_S^3$ can be as low as 0.02 and 0.06, respectively.
And from Fig. \ref{eps:sighrd},
we find that the LHC can generally detect the coupling $\la_S$ down to lower than 1.0
for $m_{\phi}$ less than 1.4 TeV.
For $m_{\phi}$ larger than 1.4 TeV, the coupling $\la_S$ needed  to discovery the monotop signal
increases quickly.
The increase of the integrated luminosity has a larger impact for larger $m_{\phi}$.
Moreover, the narrow bands of the lines, which correspond to the value
of $m_{\chi}$ varying from 5 GeV to 100 GeV, indicate the weak dependence
of the discovery potential on the value of $m_{\chi}$ if $m_{\chi}\ll m_{\phi}$.

\begin{figure}
  % Requires \usepackage{graphicx}
  \includegraphics[width=0.5\linewidth]{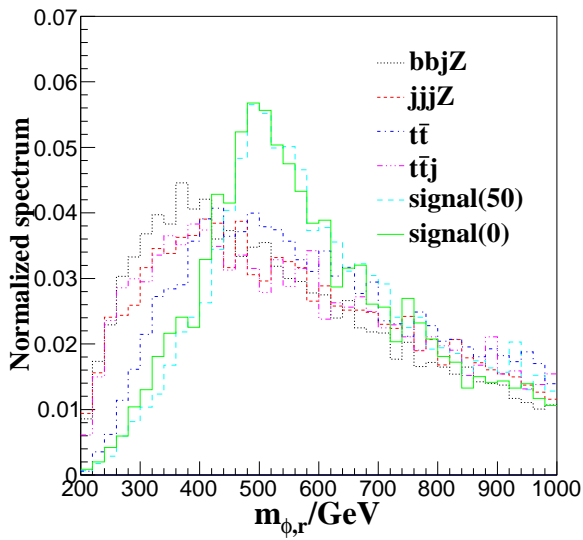}\\
  \caption{The reconstructed $m_{\phi}$ distribution for the signal and backgrounds processes.
  The signal(50) and signal(0) represent that the values of $m_{\chi}$ in Eq. (\ref{eqs:mphi}) are
  50 GeV and 0 GeV, respectively.}
  \label{eps:mphi}
\end{figure}

In this mode, since the full kinematic information of the top quark can be reconstructed,
the mass of the resonance $\phi$ can be obtained by
\begin{equation}\label{eqs:mphi}
    m_{\phi}=\sqrt{p_t^2+m_{\chi}^2}+\sqrt{p_t^2+m_t^2},
\end{equation}
with
\begin{equation}\label{eqs:pt2}
    p_t^2=p_{t,x}^2+p_{t,y}^2+p_{t,z}^2,
\end{equation}
in which $p_{t,x}, p_{t,y}, p_{t,z}$ are the three-vector momentum of the top quark.
Fig. \ref{eps:mphi} shows the distribution of the reconstructed $m_{\phi}$.
We can see a peak around $m_{\phi}=500$ GeV in the signal.
To illustrate the effect of $m_{\chi}$, we also plot the situation that
$m_{\chi}=0$ GeV is assumed in Eq. (\ref{eqs:mphi}) when reconstructing $m_{\phi}$.
It is evident that the peak position does not changed.
This information, combined with the missing transverse energy distribution,
may help to specify the masses of the resonance and the missing particle.

\begin{table}
  \centering
  \begin{tabular}{lcccccc}
  \hline\hline
  % after \\: \hline or \cline{col1-col2} \cline{col3-col4} ...
  $\sigma$ (fb) & basic & $\Slash{E}_T$ & $m_{t,r}$  & b-tagging & $\ep_{cut}$ \\
  \hline\hline
  signal            & 902               & 811               & 502              & 251  & 27.1$\%$ \\
  $jjjZ(\nu \bar{\nu})$       & $7.03\times 10^4$ & $7.87\times 10^3$ & 944              & 9.35 & 0.013$\%$ \\
  $b\bar{b}jZ(\nu \bar{\nu})$ & $1.70\times 10^3$ & 143               & 19.4             & 9.67 & 0.57$\%$ \\
  $t\bar{t}                 $ & $2.80\times 10^4$ & $34.6$            & 0.28             & 0.14 & 5 $\times 10^{-6}$ \\
  $t(\bar{t})j              $ & $2.35\times 10^4$ & $10.9$            & 0.24             & 0.12 & 5 $\times 10^{-6}$ \\
  \hline
  \end{tabular}
  \caption{The cross sections of the signal and backgrounds  after various cuts in the hadronic mode
  at the LHC ($\sqrt{s}=$ 7 TeV).
  The cut acceptance $\ep_{cut}$ is also listed.
  The entries after the $m_{t,r}$ cut for $t\bar{t} $ and $t(\bar{t})j$ processes
  are estimated by considering that one out of the total events
  we have generated for analysis can survive various kinematic cuts.}
  \label{tab:hadronic}
\end{table}

\subsection{Semileptonic mode}
For the semileptonic mode, the dominant backgrounds are $pp\to W(l \nu)j$
with the jet misidentified as a $b$-jet and single top production
with semileptonic top quark decay.
The $Wj$ background is very large
because there are only two final-state particles, compared with four final-state particles
in $pp\to jjjZ$ and $pp \to b\bar{b}jZ$ processes.
Besides, the final state of the signal contains two missing particles,
which makes the reconstruction of the mass of the top quark
very challenging.
Nevertheless, the semileptonic mode is still promising once appropriate cuts are imposed.
The signal and backgrounds are simulated by MadGraph5v1.3.3 \cite{Alwall:2011uj}
 interfaced with PYTHIA \cite{Sjostrand:2007gs}.
We choose the same default parameters as in hadronic mode, and the basic cuts are
\begin{eqnarray}
% \nonumber to remove numbering (before each equation)
  p_T^{b} > 30{~\rm GeV},\quad  |\eta^{b,l}| < 2.4, \quad {\De R}_{bl} > 0.5.
\end{eqnarray}\label{eqs:cuts_lep}

\begin{figure}
  % Requires \usepackage{graphicx}
  \includegraphics[width=0.5\linewidth]{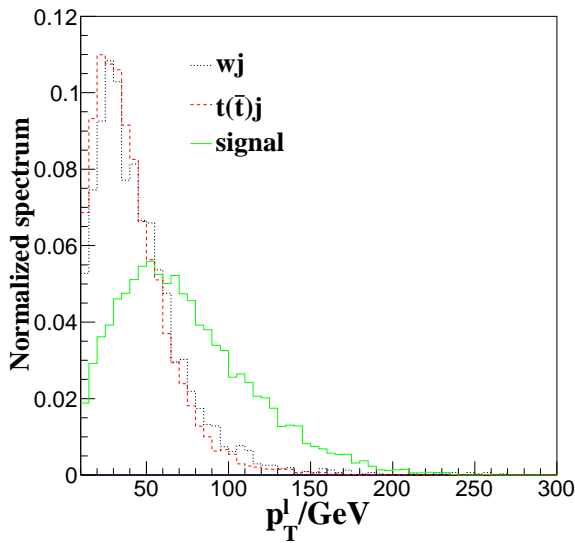}   \\
  \caption{The normalized spectrum of transverse momentum of the charged lepton in the semileptonic mode
  at the LHC ($\sqrt{s}=$ 7 TeV).}
  \label{eps:lpt}
\end{figure}

Fig. \ref{eps:lpt} shows the normalized spectrum of the transverse momentum of the charged lepton
in the semileptonic mode at the LHC with $\sqrt{s}=$ 7 TeV. We can see that
it is difficult to suppress the backgrounds by $p_T^{l}$ cut because of the similar distributions of the
signal and backgrounds. As a result, we choose a loose cut
\begin{equation}
    p_T^{l} > 20{~\rm GeV}
\end{equation}
to keep more signal events.

\begin{figure}
  % Requires \usepackage{graphicx}
  \includegraphics[width=0.5\linewidth]{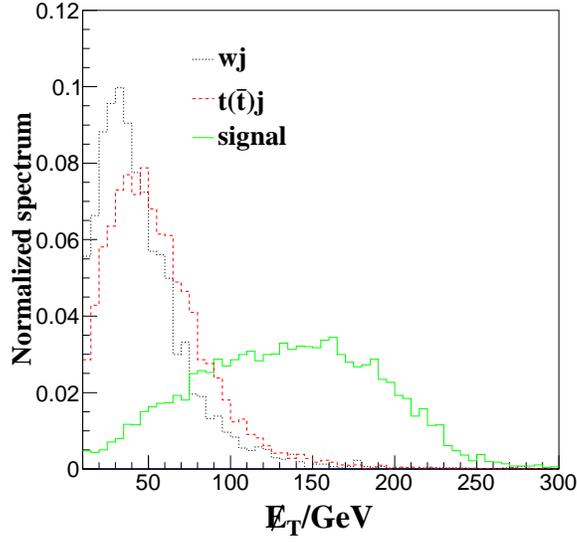}  \\
  \caption{The normalized spectrum of the missing transverse energy in the semileptonic mode
  at the LHC ($\sqrt{s}=$ 7 TeV).}
  \label{eps:mpt_l}
\end{figure}

Fig. \ref{eps:mpt_l} shows the normalized spectrum of the missing transverse energy
in the semileptonic mode at the LHC with $\sqrt{s}=$ 7 TeV.
The backgrounds decrease while the signal increases in the range $30{~\rm GeV}<\Slash{E}_T <150{~\rm GeV}$.
The reason is that the missing particle of the backgrounds is (anti)neutrino,
which comes from the $W$ boson, and the $Wj$ is mainly produced through t-channel,
in which the momentum of final-state particles tend to be collinear to those of the initial-state particles.
The situation for the single top production is similar.
In contrast, the missing particles of the signal originate from a resonance of a large mass, and thus
could be produced with large transverse momentum.
Therefore, we impose the missing transverse energy cut
\begin{equation}
    \Slash{E}_T > 120{~\rm GeV}
\end{equation}
to suppress the backgrounds.

\begin{figure}
  % Requires \usepackage{graphicx}
  \includegraphics[width=0.5\linewidth]{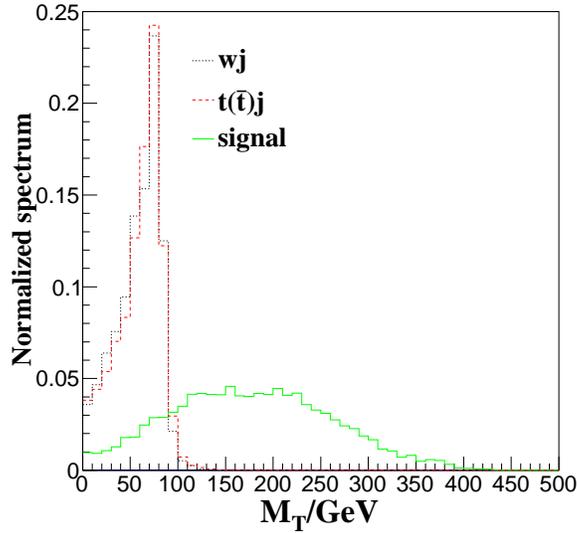}  \\
  \caption{The normalized spectrum of the transverse mass $M_{T}$ in the semileptonic mode
  at the LHC ($\sqrt{s}=$ 7 TeV).}
  \label{eps:tmpt_l}
\end{figure}

Fig. \ref{eps:tmpt_l} shows the normalized spectrum of the transverse mass,
which is defined as \cite{Nakamura:2010zzi}
\begin{equation}\label{eq:tranmass}
    M_T=\sqrt{(\Slash{E}_T+E_T^l)^2-(\vec{\Slash{p}}_T+\vec{p_T^l})^2},
\end{equation}
in the semileptonic mode at the LHC with $\sqrt{s}=$ 7 TeV.
The backgrounds increase in the range $0<M_T <80{~\rm GeV}$ and have a peak around $M_T\sim 80{~\rm GeV}$.
This is due to the fact that the transverse mass measure the maximum of the invariant mass of
the missing particles and the lepton, which is the mass of $W$ boson for the backgrounds.
In contrast, the signal concentrates in the range $M_T >100{~\rm GeV}$.
Thus, to suppress the backgrounds efficiently, we impose the transverse mass cut
\begin{equation}
     M_T > 120{~\rm GeV}.
\end{equation}

\begin{figure}
  % Requires \usepackage{graphicx}
  \includegraphics[width=0.5\linewidth]{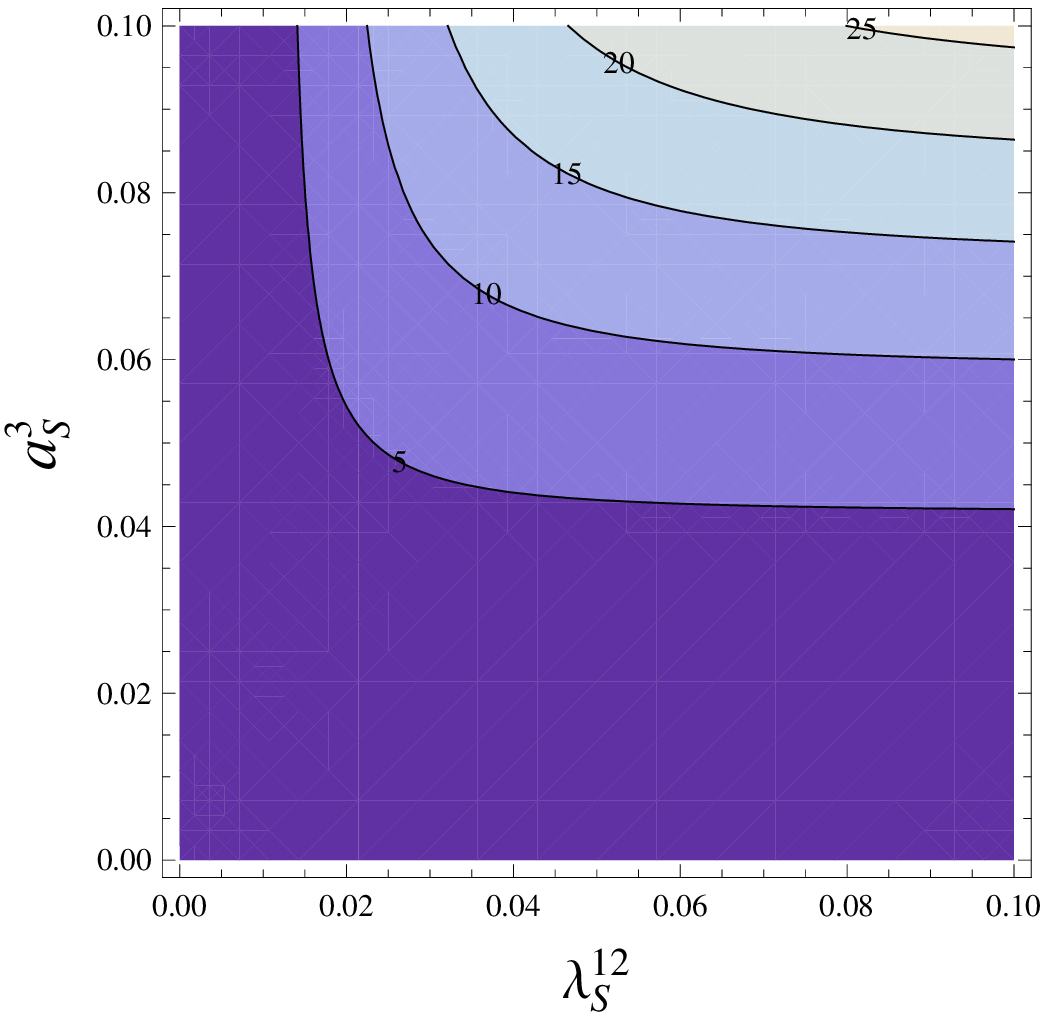}\\
  \caption{The significance in the semileptonic mode at the LHC ($\sqrt{s}=$ 7 TeV)
  with an integrated luminosity of 1 $\rm fb^{-1}$
   versus the parameters $\la_S^{12}$ and $a_S^3$, assuming $m_{\phi}=500{~\rm GeV} {\rm ~and~} m_{\chi}=50{~\rm GeV}$. }
  \label{eps:SignifLep}
\end{figure}

\begin{figure}
  % Requires \usepackage{graphicx}
  \includegraphics[width=0.5\linewidth]{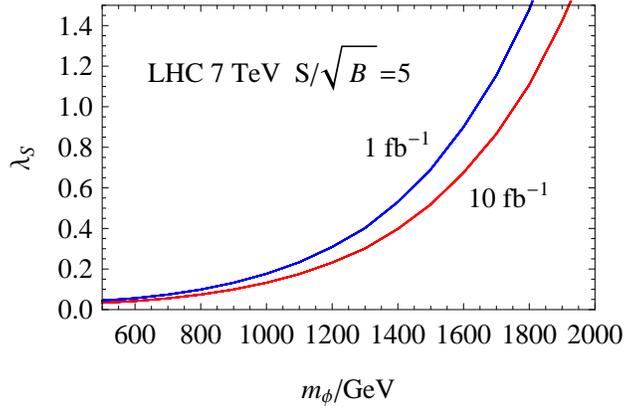}\\
  \caption{The $5\sigma$ discovery limits of $m_{\phi}$ and $\la_S(=\la_S^{12}=a_S^3)$
  in the semileptonic mode at the LHC ($\sqrt{s}=$ 7 TeV).
  Either band consists of twenty solid lines from the bottom up
  corresponding to the value of $m_{\chi}$ varying from 5 GeV to 100 GeV with a step of 5 GeV. }
  \label{eps:LamLep}
\end{figure}

The cross sections of the signal and backgrounds after various cuts at the LHC  ($\sqrt{s}=$ 7 TeV)
 are listed in Table~\ref{tab:leptonic}.
We can see that the backgrounds nearly vanish after the transverse mass cut is imposed,
which means that it is very promising to search for the signal of monotops in the semileptonic mode.
In Fig. \ref{eps:SignifLep}, we show the contour curves of the significance $\mathcal{S}$
versus the parameters $\la_S^{12}$ and $a_S^3$ in the semileptonic mode at the LHC ($\sqrt{s}=$ 7 TeV).
And in Fig. \ref{eps:LamLep}, we show the $5\sigma$ ($\mathcal{S}=5$) discovery limits of
$m_{\phi}$, $m_{\chi}$ and $\la_S^{12}=a_S^3=\la_S$ in the semileptonic mode.
From Fig. \ref{eps:SignifLep} we can see that for a $5\sigma$ discovery,
the sensitivity to $\la_S^{12}$ and $a_S^3$ can be as low as 0.015 and 0.045, respectively,
which are smaller than the corresponding values in the hadronic mode.
And from Fig. \ref{eps:LamLep},
we find that the LHC can generally detect the coupling $\la_S$ down to lower than 0.4
for $m_{\phi}$ less than 1.4 TeV, and
for larger $m_{\phi}$, the coupling $\la_S$ needed  to discover the monotop signal
increases quickly.
Also, the value of $m_{\chi}$ has little effect on the discovery potential.

\begin{table}
  \centering
  \begin{tabular}{lcccccc}
  \hline\hline
  % after \\: \hline or \cline{col1-col2} \cline{col3-col4} ...
  $\sigma $ (fb)& basic & $p_T^l$ & $\Slash{E}_T$ & $M_T$ & b-tagging & $\ep_{cut}$ \\
  \hline\hline
  signal & 399               & 376               & 231               & 218  & 109 & 27.3$\%$ \\
  $W(l \nu)j$     & $1.83\times 10^6$ & $1.53\times 10^6$ & $3.45\times 10^4$ & $1.83 $ & 0.003   & 2$\times 10^{-9}$  \\
  $t(\bar{t})j$     & $9.09\times 10^3$ & $7.33\times 10^3$ & $185$       & $2.15 $ & 1.08   & $0.012\%$  \\
  \hline
  \end{tabular}
  \caption{The cross sections of the signal and backgrounds after various cuts in the semileptonic mode
  at the LHC ($\sqrt{s}=$ 7 TeV).
  The cut acceptance $\ep_{cut}$ is also listed.
    The entries after the $M_T$ cut for  $W(l \nu)j$ process
  are estimated by considering that one out of the total events
  we have generated for analysis can survive various kinematic cuts.}
  \label{tab:leptonic}
\end{table}

\section{Conclusion}\label{sec:conclusion}

We have investigated the potential of the early LHC to discover the signal of monotop production.
First, we obtain the parameter space of the effective Lagrangian constrained by the present data of
$Z$ boson hadronic decay branching ratio, $K^0-\overline{K^0}$ mixing and dijet productions at the LHC.
Then, we study the various cuts imposed on the events, reconstructed from the hadronic final states,
to suppress backgrounds and increase the significance in detail.
And we find that in the hadronic mode the information from the missing transverse energy and reconstructed resonance mass
distributions can be used to specify the masses of the resonance and the missing particle.
Lastly, we present the significance $\mathcal{S}$  at the LHC ($\sqrt{s}=$ 7 TeV) with an integrated luminosity
of 1 $\rm fb^{-1}$ in the parameter space allowed by the current data, and  the
$5\sigma$ discovery limits of $m_{\phi}$ and $\la_S(=\la_S^{12}=a_S^3)$.
Our results show that the LHC can generally detect the coupling $\la_S$
down to lower than 1.0 and 0.4 for $m_{\phi}$ less than 1.4 TeV
in the hadronic and semileptonic modes, respectively.

\acknowledgments
This work was supported by the National Natural
Science Foundation of China, under Grants
No. 11021092, No. 10975004 and No. 11135003.

\newpage %Just because of unusual number of tables stacked at end
\bibliography{monotop}% Produces the bibliography via BibTeX.

\begin{thebibliography}{41}
\expandafter\ifx\csname natexlab\endcsname\relax\def\natexlab#1{#1}\fi
\expandafter\ifx\csname bibnamefont\endcsname\relax
  \def\bibnamefont#1{#1}\fi
\expandafter\ifx\csname bibfnamefont\endcsname\relax
  \def\bibfnamefont#1{#1}\fi
\expandafter\ifx\csname citenamefont\endcsname\relax
  \def\citenamefont#1{#1}\fi
\expandafter\ifx\csname url\endcsname\relax
  \def\url#1{\texttt{#1}}\fi
\expandafter\ifx\csname urlprefix\endcsname\relax\def\urlprefix{URL }\fi
\providecommand{\bibinfo}[2]{#2}
\providecommand{\eprint}[2][]{\url{#2}}

\bibitem[{\citenamefont{Alvarez et~al.}(2011)\citenamefont{Alvarez, Da~Rold,
  Vietto, and Szynkman}}]{Alvarez:2011hi}
\bibinfo{author}{\bibfnamefont{E.}~\bibnamefont{Alvarez}},
  \bibinfo{author}{\bibfnamefont{L.}~\bibnamefont{Da~Rold}},
  \bibinfo{author}{\bibfnamefont{J.~I.~S.} \bibnamefont{Vietto}},
  \bibnamefont{and} \bibinfo{author}{\bibfnamefont{A.}~\bibnamefont{Szynkman}},
  \bibinfo{journal}{JHEP} \textbf{\bibinfo{volume}{1109}}, \bibinfo{pages}{007}
  (\bibinfo{year}{2011}), \eprint{1107.1473}.

\bibitem[{\citenamefont{Haisch and Westhoff}(2011)}]{Haisch:2011up}
\bibinfo{author}{\bibfnamefont{U.}~\bibnamefont{Haisch}} \bibnamefont{and}
  \bibinfo{author}{\bibfnamefont{S.}~\bibnamefont{Westhoff}}
  (\bibinfo{year}{2011}), \bibinfo{note}{* Temporary entry *},
  \eprint{1106.0529}.

\bibitem[{\citenamefont{Berger et~al.}(2011)\citenamefont{Berger, Cao, Chen,
  and Zhang}}]{Berger:2011hn}
\bibinfo{author}{\bibfnamefont{E.~L.} \bibnamefont{Berger}},
  \bibinfo{author}{\bibfnamefont{Q.-H.} \bibnamefont{Cao}},
  \bibinfo{author}{\bibfnamefont{C.-R.} \bibnamefont{Chen}}, \bibnamefont{and}
  \bibinfo{author}{\bibfnamefont{H.}~\bibnamefont{Zhang}},
  \bibinfo{journal}{Phys.Rev.} \textbf{\bibinfo{volume}{D83}},
  \bibinfo{pages}{114026} (\bibinfo{year}{2011}), \eprint{1103.3274}.

\bibitem[{\citenamefont{Cao et~al.}(2011)\citenamefont{Cao, Wu, and
  Yang}}]{Cao:2010nw}
\bibinfo{author}{\bibfnamefont{J.}~\bibnamefont{Cao}},
  \bibinfo{author}{\bibfnamefont{L.}~\bibnamefont{Wu}}, \bibnamefont{and}
  \bibinfo{author}{\bibfnamefont{J.~M.} \bibnamefont{Yang}},
  \bibinfo{journal}{Phys.Rev.} \textbf{\bibinfo{volume}{D83}},
  \bibinfo{pages}{034024} (\bibinfo{year}{2011}), \eprint{1011.5564}.

\bibitem[{\citenamefont{Degrande et~al.}(2011)\citenamefont{Degrande, Gerard,
  Grojean, Maltoni, and Servant}}]{Degrande:2010kt}
\bibinfo{author}{\bibfnamefont{C.}~\bibnamefont{Degrande}},
  \bibinfo{author}{\bibfnamefont{J.-M.} \bibnamefont{Gerard}},
  \bibinfo{author}{\bibfnamefont{C.}~\bibnamefont{Grojean}},
  \bibinfo{author}{\bibfnamefont{F.}~\bibnamefont{Maltoni}}, \bibnamefont{and}
  \bibinfo{author}{\bibfnamefont{G.}~\bibnamefont{Servant}},
  \bibinfo{journal}{JHEP} \textbf{\bibinfo{volume}{03}}, \bibinfo{pages}{125}
  (\bibinfo{year}{2011}), \eprint{1010.6304}.

\bibitem[{\citenamefont{Battaglia and Servant}(2010)}]{Battaglia:2010xq}
\bibinfo{author}{\bibfnamefont{M.}~\bibnamefont{Battaglia}} \bibnamefont{and}
  \bibinfo{author}{\bibfnamefont{G.}~\bibnamefont{Servant}}
  (\bibinfo{year}{2010}), \eprint{1005.4632}.

\bibitem[{\citenamefont{Cao et~al.}(2010)\citenamefont{Cao, McKeen, Rosner,
  Shaughnessy, and Wagner}}]{Cao:2010zb}
\bibinfo{author}{\bibfnamefont{Q.-H.} \bibnamefont{Cao}},
  \bibinfo{author}{\bibfnamefont{D.}~\bibnamefont{McKeen}},
  \bibinfo{author}{\bibfnamefont{J.~L.} \bibnamefont{Rosner}},
  \bibinfo{author}{\bibfnamefont{G.}~\bibnamefont{Shaughnessy}},
  \bibnamefont{and} \bibinfo{author}{\bibfnamefont{C.~E.M.}
  \bibnamefont{Wagner}}, \bibinfo{journal}{Phys.Rev.}
  \textbf{\bibinfo{volume}{D81}}, \bibinfo{pages}{114004}
  (\bibinfo{year}{2010}), \eprint{1003.3461}.

\bibitem[{\citenamefont{Alwall et~al.}(2010)\citenamefont{Alwall, Feng, Kumar,
  and Su}}]{Alwall:2010jc}
\bibinfo{author}{\bibfnamefont{J.}~\bibnamefont{Alwall}},
  \bibinfo{author}{\bibfnamefont{J.~L.} \bibnamefont{Feng}},
  \bibinfo{author}{\bibfnamefont{J.}~\bibnamefont{Kumar}}, \bibnamefont{and}
  \bibinfo{author}{\bibfnamefont{S.}~\bibnamefont{Su}}, \bibinfo{journal}{Phys.
  Rev.} \textbf{\bibinfo{volume}{D81}}, \bibinfo{pages}{114027}
  (\bibinfo{year}{2010}), \eprint{1002.3366}.

\bibitem[{\citenamefont{Han et~al.}(2009)\citenamefont{Han, Mahbubani, Walker,
  and Wang}}]{Han:2008gy}
\bibinfo{author}{\bibfnamefont{T.}~\bibnamefont{Han}},
  \bibinfo{author}{\bibfnamefont{R.}~\bibnamefont{Mahbubani}},
  \bibinfo{author}{\bibfnamefont{D.~G.} \bibnamefont{Walker}},
  \bibnamefont{and} \bibinfo{author}{\bibfnamefont{L.-T.} \bibnamefont{Wang}},
  \bibinfo{journal}{JHEP} \textbf{\bibinfo{volume}{0905}}, \bibinfo{pages}{117}
  (\bibinfo{year}{2009}), \eprint{0803.3820}.

\bibitem[{\citenamefont{Barger et~al.}(2008)\citenamefont{Barger, Han, and
  Walker}}]{Barger:2006hm}
\bibinfo{author}{\bibfnamefont{V.}~\bibnamefont{Barger}},
  \bibinfo{author}{\bibfnamefont{T.}~\bibnamefont{Han}}, \bibnamefont{and}
  \bibinfo{author}{\bibfnamefont{D.~G.E} \bibnamefont{Walker}},
  \bibinfo{journal}{Phys.Rev.Lett.} \textbf{\bibinfo{volume}{100}},
  \bibinfo{pages}{031801} (\bibinfo{year}{2008}), \eprint{hep-ph/0612016}.

\bibitem[{\citenamefont{Andrea et~al.}(2011)\citenamefont{Andrea, Fuks, and
  Maltoni}}]{Andrea:2011ws}
\bibinfo{author}{\bibfnamefont{J.}~\bibnamefont{Andrea}},
  \bibinfo{author}{\bibfnamefont{B.}~\bibnamefont{Fuks}}, \bibnamefont{and}
  \bibinfo{author}{\bibfnamefont{F.}~\bibnamefont{Maltoni}}
  (\bibinfo{year}{2011}), \eprint{1106.6199}.

\bibitem[{\citenamefont{Kamenik and Zupan}(2011)}]{Kamenik:2011nb}
\bibinfo{author}{\bibfnamefont{J.~F.} \bibnamefont{Kamenik}} \bibnamefont{and}
  \bibinfo{author}{\bibfnamefont{J.}~\bibnamefont{Zupan}}
  (\bibinfo{year}{2011}), \eprint{1107.0623}.

\bibitem[{\citenamefont{Dong et~al.}(2011)\citenamefont{Dong, Durieux, Gerard,
  Han, and Maltoni}}]{Dong:2011rh}
\bibinfo{author}{\bibfnamefont{Z.}~\bibnamefont{Dong}},
  \bibinfo{author}{\bibfnamefont{G.}~\bibnamefont{Durieux}},
  \bibinfo{author}{\bibfnamefont{J.-M.} \bibnamefont{Gerard}},
  \bibinfo{author}{\bibfnamefont{T.}~\bibnamefont{Han}}, \bibnamefont{and}
  \bibinfo{author}{\bibfnamefont{F.}~\bibnamefont{Maltoni}}
  (\bibinfo{year}{2011}), \bibinfo{note}{* Temporary entry *},
  \eprint{1107.3805}.

\bibitem[{\citenamefont{Barbier et~al.}(2005)\citenamefont{Barbier, Berat,
  Besancon, Chemtob, Deandrea et~al.}}]{Barbier:2004ez}
\bibinfo{author}{\bibfnamefont{R.}~\bibnamefont{Barbier}},
  \bibinfo{author}{\bibfnamefont{C.}~\bibnamefont{Berat}},
  \bibinfo{author}{\bibfnamefont{M.}~\bibnamefont{Besancon}},
  \bibinfo{author}{\bibfnamefont{M.}~\bibnamefont{Chemtob}},
  \bibinfo{author}{\bibfnamefont{A.}~\bibnamefont{Deandrea}},
  \bibnamefont{et~al.}, \bibinfo{journal}{Phys.Rept.}
  \textbf{\bibinfo{volume}{420}}, \bibinfo{pages}{1} (\bibinfo{year}{2005}),
  \eprint{hep-ph/0406039}.

\bibitem[{\citenamefont{Barr}(1982)}]{Barr:1981qv}
\bibinfo{author}{\bibfnamefont{S.~M.} \bibnamefont{Barr}},
  \bibinfo{journal}{Phys.Lett.} \textbf{\bibinfo{volume}{B112}},
  \bibinfo{pages}{219} (\bibinfo{year}{1982}).

\bibitem[{\citenamefont{Aktas et~al.}(2004)}]{Aktas:2004ij}
\bibinfo{author}{\bibfnamefont{A.}~\bibnamefont{Aktas}} \bibnamefont{et~al.}
  (\bibinfo{collaboration}{H1}), \bibinfo{journal}{Eur. Phys. J.}
  \textbf{\bibinfo{volume}{C36}}, \bibinfo{pages}{425} (\bibinfo{year}{2004}),
  \eprint{hep-ex/0403027}.

\bibitem[{\citenamefont{Chekanov et~al.}(2007)}]{:2006je}
\bibinfo{author}{\bibfnamefont{S.}~\bibnamefont{Chekanov}} \bibnamefont{et~al.}
  (\bibinfo{collaboration}{ZEUS}), \bibinfo{journal}{Eur. Phys. J.}
  \textbf{\bibinfo{volume}{C50}}, \bibinfo{pages}{269} (\bibinfo{year}{2007}),
  \eprint{hep-ex/0611018}.

\bibitem[{\citenamefont{Chakrabarti et~al.}(2003)\citenamefont{Chakrabarti,
  Guchait, and Mondal}}]{Chakrabarti:2003wr}
\bibinfo{author}{\bibfnamefont{S.}~\bibnamefont{Chakrabarti}},
  \bibinfo{author}{\bibfnamefont{M.}~\bibnamefont{Guchait}}, \bibnamefont{and}
  \bibinfo{author}{\bibfnamefont{N.~K.} \bibnamefont{Mondal}},
  \bibinfo{journal}{Phys. Rev.} \textbf{\bibinfo{volume}{D68}},
  \bibinfo{pages}{015005} (\bibinfo{year}{2003}), \eprint{hep-ph/0301248}.

\bibitem[{\citenamefont{Bhattacharyya et~al.}(1995)\citenamefont{Bhattacharyya,
  Choudhury, and Sridhar}}]{Bhattacharyya:1995bw}
\bibinfo{author}{\bibfnamefont{G.}~\bibnamefont{Bhattacharyya}},
  \bibinfo{author}{\bibfnamefont{D.}~\bibnamefont{Choudhury}},
  \bibnamefont{and} \bibinfo{author}{\bibfnamefont{K.}~\bibnamefont{Sridhar}},
  \bibinfo{journal}{Phys. Lett.} \textbf{\bibinfo{volume}{B355}},
  \bibinfo{pages}{193} (\bibinfo{year}{1995}), \eprint{hep-ph/9504314}.

\bibitem[{\citenamefont{Nakamura et~al.}(2010)}]{Nakamura:2010zzi}
\bibinfo{author}{\bibfnamefont{K.}~\bibnamefont{Nakamura}} \bibnamefont{et~al.}
  (\bibinfo{collaboration}{Particle Data Group}), \bibinfo{journal}{J.Phys.G}
  \textbf{\bibinfo{volume}{G37}}, \bibinfo{pages}{075021}
  (\bibinfo{year}{2010}).

\bibitem[{\citenamefont{Herrlich and Nierste}(1995)}]{Herrlich:1995hh}
\bibinfo{author}{\bibfnamefont{S.}~\bibnamefont{Herrlich}} \bibnamefont{and}
  \bibinfo{author}{\bibfnamefont{U.}~\bibnamefont{Nierste}},
  \bibinfo{journal}{Phys. Rev.} \textbf{\bibinfo{volume}{D52}},
  \bibinfo{pages}{6505} (\bibinfo{year}{1995}), \eprint{hep-ph/9507262}.

\bibitem[{\citenamefont{de~Carlos and White}(1997)}]{deCarlos:1996yh}
\bibinfo{author}{\bibfnamefont{B.}~\bibnamefont{de~Carlos}} \bibnamefont{and}
  \bibinfo{author}{\bibfnamefont{P.~L.} \bibnamefont{White}},
  \bibinfo{journal}{Phys. Rev.} \textbf{\bibinfo{volume}{D55}},
  \bibinfo{pages}{4222} (\bibinfo{year}{1997}), \eprint{hep-ph/9609443}.

\bibitem[{\citenamefont{Slavich}(2001)}]{Slavich:2000xm}
\bibinfo{author}{\bibfnamefont{P.}~\bibnamefont{Slavich}},
  \bibinfo{journal}{Nucl. Phys.} \textbf{\bibinfo{volume}{B595}},
  \bibinfo{pages}{33} (\bibinfo{year}{2001}), \eprint{hep-ph/0008270}.

\bibitem[{\citenamefont{Ciuchini et~al.}(1998)\citenamefont{Ciuchini, Lubicz,
  Conti, Vladikas, Donini et~al.}}]{Ciuchini:1998ix}
\bibinfo{author}{\bibfnamefont{M.}~\bibnamefont{Ciuchini}},
  \bibinfo{author}{\bibfnamefont{V.}~\bibnamefont{Lubicz}},
  \bibinfo{author}{\bibfnamefont{L.}~\bibnamefont{Conti}},
  \bibinfo{author}{\bibfnamefont{A.}~\bibnamefont{Vladikas}},
  \bibinfo{author}{\bibfnamefont{A.}~\bibnamefont{Donini}},
  \bibnamefont{et~al.}, \bibinfo{journal}{JHEP}
  \textbf{\bibinfo{volume}{9810}}, \bibinfo{pages}{008} (\bibinfo{year}{1998}),
  \bibinfo{note}{erratum added online, Mar/29/2000}, \eprint{hep-ph/9808328}.

\bibitem[{\citenamefont{Buras}(1998)}]{Buras:1998raa}
\bibinfo{author}{\bibfnamefont{A.~J.} \bibnamefont{Buras}}
  (\bibinfo{year}{1998}), \eprint{hep-ph/9806471}.

\bibitem[{\citenamefont{Buras et~al.}(1990)\citenamefont{Buras, Jamin, and
  Weisz}}]{Buras:1990fn}
\bibinfo{author}{\bibfnamefont{A.~J.} \bibnamefont{Buras}},
  \bibinfo{author}{\bibfnamefont{M.}~\bibnamefont{Jamin}}, \bibnamefont{and}
  \bibinfo{author}{\bibfnamefont{P.~H.} \bibnamefont{Weisz}},
  \bibinfo{journal}{Nucl. Phys.} \textbf{\bibinfo{volume}{B347}},
  \bibinfo{pages}{491} (\bibinfo{year}{1990}).

\bibitem[{\citenamefont{Herrlich and Nierste}(1994)}]{Herrlich:1993yv}
\bibinfo{author}{\bibfnamefont{S.}~\bibnamefont{Herrlich}} \bibnamefont{and}
  \bibinfo{author}{\bibfnamefont{U.}~\bibnamefont{Nierste}},
  \bibinfo{journal}{Nucl. Phys.} \textbf{\bibinfo{volume}{B419}},
  \bibinfo{pages}{292} (\bibinfo{year}{1994}), \eprint{hep-ph/9310311}.

\bibitem[{\citenamefont{Urban et~al.}(1998)\citenamefont{Urban, Krauss,
  Jentschura, and Soff}}]{Urban:1997gw}
\bibinfo{author}{\bibfnamefont{J.}~\bibnamefont{Urban}},
  \bibinfo{author}{\bibfnamefont{F.}~\bibnamefont{Krauss}},
  \bibinfo{author}{\bibfnamefont{U.}~\bibnamefont{Jentschura}},
  \bibnamefont{and} \bibinfo{author}{\bibfnamefont{G.}~\bibnamefont{Soff}},
  \bibinfo{journal}{Nucl. Phys.} \textbf{\bibinfo{volume}{B523}},
  \bibinfo{pages}{40} (\bibinfo{year}{1998}), \eprint{hep-ph/9710245}.

\bibitem[{\citenamefont{Khachatryan et~al.}(2010)}]{Khachatryan:2010jd}
\bibinfo{author}{\bibfnamefont{V.}~\bibnamefont{Khachatryan}}
  \bibnamefont{et~al.} (\bibinfo{collaboration}{CMS}), \bibinfo{journal}{Phys.
  Rev. Lett.} \textbf{\bibinfo{volume}{105}}, \bibinfo{pages}{211801}
  (\bibinfo{year}{2010}), \eprint{1010.0203}.

\bibitem[{\citenamefont{Aad et~al.}(2010)}]{:2010bc}
\bibinfo{author}{\bibfnamefont{G.}~\bibnamefont{Aad}} \bibnamefont{et~al.}
  (\bibinfo{collaboration}{ATLAS}), \bibinfo{journal}{Phys. Rev. Lett.}
  \textbf{\bibinfo{volume}{105}}, \bibinfo{pages}{161801}
  (\bibinfo{year}{2010}), \eprint{1008.2461}.

\bibitem[{\citenamefont{Aad et~al.}(2011{\natexlab{a}})}]{Aad:2011aj}
\bibinfo{author}{\bibfnamefont{G.}~\bibnamefont{Aad}} \bibnamefont{et~al.}
  (\bibinfo{collaboration}{ATLAS}), \bibinfo{journal}{New J. Phys.}
  \textbf{\bibinfo{volume}{13}}, \bibinfo{pages}{053044}
  (\bibinfo{year}{2011}{\natexlab{a}}), \eprint{1103.3864}.

\bibitem[{\citenamefont{Chatrchyan et~al.}(2011)}]{Chatrchyan:2011ns}
\bibinfo{author}{\bibfnamefont{S.}~\bibnamefont{Chatrchyan}}
  \bibnamefont{et~al.} (\bibinfo{collaboration}{CMS}) (\bibinfo{year}{2011}),
  \eprint{1107.4771}.

\bibitem[{\citenamefont{Aad et~al.}(2011{\natexlab{b}})}]{Aad:2011fq}
\bibinfo{author}{\bibfnamefont{G.}~\bibnamefont{Aad}} \bibnamefont{et~al.}
  (\bibinfo{collaboration}{ATLAS Collaboration})
  (\bibinfo{year}{2011}{\natexlab{b}}), \eprint{1108.6311}.

\bibitem[{\citenamefont{Alwall et~al.}(2011)\citenamefont{Alwall, Herquet,
  Maltoni, Mattelaer, and Stelzer}}]{Alwall:2011uj}
\bibinfo{author}{\bibfnamefont{J.}~\bibnamefont{Alwall}},
  \bibinfo{author}{\bibfnamefont{M.}~\bibnamefont{Herquet}},
  \bibinfo{author}{\bibfnamefont{F.}~\bibnamefont{Maltoni}},
  \bibinfo{author}{\bibfnamefont{O.}~\bibnamefont{Mattelaer}},
  \bibnamefont{and} \bibinfo{author}{\bibfnamefont{T.}~\bibnamefont{Stelzer}},
  \bibinfo{journal}{JHEP} \textbf{\bibinfo{volume}{06}}, \bibinfo{pages}{128}
  (\bibinfo{year}{2011}), \eprint{1106.0522}.

\bibitem[{\citenamefont{Christensen and Duhr}(2009)}]{Christensen:2008py}
\bibinfo{author}{\bibfnamefont{N.~D.} \bibnamefont{Christensen}}
  \bibnamefont{and} \bibinfo{author}{\bibfnamefont{C.}~\bibnamefont{Duhr}},
  \bibinfo{journal}{Comput. Phys. Commun.} \textbf{\bibinfo{volume}{180}},
  \bibinfo{pages}{1614} (\bibinfo{year}{2009}), \eprint{0806.4194}.

\bibitem[{\citenamefont{Belanger et~al.}(2004)\citenamefont{Belanger, Boudjema,
  Cottrant, Pukhov, and Rosier-Lees}}]{Belanger:2003wb}
\bibinfo{author}{\bibfnamefont{G.}~\bibnamefont{Belanger}},
  \bibinfo{author}{\bibfnamefont{F.}~\bibnamefont{Boudjema}},
  \bibinfo{author}{\bibfnamefont{A.}~\bibnamefont{Cottrant}},
  \bibinfo{author}{\bibfnamefont{A.}~\bibnamefont{Pukhov}}, \bibnamefont{and}
  \bibinfo{author}{\bibfnamefont{S.}~\bibnamefont{Rosier-Lees}},
  \bibinfo{journal}{JHEP} \textbf{\bibinfo{volume}{0403}}, \bibinfo{pages}{012}
  (\bibinfo{year}{2004}), \eprint{hep-ph/0310037}.

\bibitem[{\citenamefont{Aad et~al.}(2012)}]{Aad:2011zb}
\bibinfo{author}{\bibfnamefont{G.}~\bibnamefont{Aad}} \bibnamefont{et~al.}
  (\bibinfo{collaboration}{ATLAS Collaboration}), \bibinfo{journal}{Phys.Lett.}
  \textbf{\bibinfo{volume}{B707}}, \bibinfo{pages}{478} (\bibinfo{year}{2012}),
  \eprint{1109.2242}.

\bibitem[{\citenamefont{Mangano et~al.}(2003)\citenamefont{Mangano, Moretti,
  Piccinini, Pittau, and Polosa}}]{Mangano:2002ea}
\bibinfo{author}{\bibfnamefont{M.~L.} \bibnamefont{Mangano}},
  \bibinfo{author}{\bibfnamefont{M.}~\bibnamefont{Moretti}},
  \bibinfo{author}{\bibfnamefont{F.}~\bibnamefont{Piccinini}},
  \bibinfo{author}{\bibfnamefont{R.}~\bibnamefont{Pittau}}, \bibnamefont{and}
  \bibinfo{author}{\bibfnamefont{A.~D.} \bibnamefont{Polosa}},
  \bibinfo{journal}{JHEP} \textbf{\bibinfo{volume}{0307}}, \bibinfo{pages}{001}
  (\bibinfo{year}{2003}), \eprint{hep-ph/0206293}.

\bibitem[{\citenamefont{Sjostrand et~al.}(2008)\citenamefont{Sjostrand, Mrenna,
  and Skands}}]{Sjostrand:2007gs}
\bibinfo{author}{\bibfnamefont{T.}~\bibnamefont{Sjostrand}},
  \bibinfo{author}{\bibfnamefont{S.}~\bibnamefont{Mrenna}}, \bibnamefont{and}
  \bibinfo{author}{\bibfnamefont{P.~Z.} \bibnamefont{Skands}},
  \bibinfo{journal}{Comput.Phys.Commun.} \textbf{\bibinfo{volume}{178}},
  \bibinfo{pages}{852} (\bibinfo{year}{2008}), \eprint{0710.3820}.

\bibitem[{\citenamefont{Sjostrand et~al.}(2006)\citenamefont{Sjostrand, Mrenna,
  and Skands}}]{Sjostrand:2006za}
\bibinfo{author}{\bibfnamefont{T.}~\bibnamefont{Sjostrand}},
  \bibinfo{author}{\bibfnamefont{S.}~\bibnamefont{Mrenna}}, \bibnamefont{and}
  \bibinfo{author}{\bibfnamefont{P.~Z.} \bibnamefont{Skands}},
  \bibinfo{journal}{JHEP} \textbf{\bibinfo{volume}{0605}}, \bibinfo{pages}{026}
  (\bibinfo{year}{2006}), \eprint{hep-ph/0603175}.

\bibitem[{\citenamefont{Aad et~al.}(2009)}]{Aad:2009wy}
\bibinfo{author}{\bibfnamefont{G.}~\bibnamefont{Aad}} \bibnamefont{et~al.}
  (\bibinfo{collaboration}{The ATLAS Collaboration}) (\bibinfo{year}{2009}),
  \eprint{0901.0512}.

\end{thebibliography}

\end{document}